\documentstyle[preprint,prl,aps,epsfig,floats]{revtex}
\begin{document}
\draft
\preprint{
\vbox{\halign{&##\hfil        \cr
        & DESY 00-067      \cr
        & hep-ph/0004228      \cr
        & April 2000          \cr
        }}}

\title{ Next-to-leading Order Calculation of the \\
	Color-Octet $^3S_1$ Gluon Fragmentation Function \\
	for Heavy Quarkonium}

\author{Eric Braaten}
\address{Physics Department, Ohio State University, Columbus OH 43210, USA}

\author{Jungil Lee}
\address{II. Institut f\"ur Theoretische Physik,
         Universit\"at Hamburg,
        22761 Hamburg, Germany}

\maketitle
\begin{abstract}
The short-distance coefficients for the color-octet $^3S_1$ term
in the fragmentation function for a gluon to split into 
heavy quarkonium states is calculated to order $\alpha_s^2$.
The gauge-invariant definition of the fragmentation function by Collins and
Soper is employed. Ultraviolet divergences are removed using the 
$\overline{\rm MS}$ renormalization procedure.
The longitudinal term in the fragmentation function agrees with 
a previous calculation by Beneke and Rothstein.
The next-to-leading order correction to the transverse term
disagrees with a previous calculation.
\end{abstract}

\vfill \eject

\narrowtext
\section{Introduction}
The cross sections for heavy quarkonium states probe 
the production of 
heavy-quark-antiquark pairs with small relative momenta.   
Many of the theoretical uncertainties
in quarkonium production decrease at large transverse momentum.
Factorization theorems for inclusive single-hadron production
\cite{C-S} guarantee that the dominant mechanism for 
producing heavy quarkonia with high $p_T$
is fragmentation\cite{B-Y:S1}, the production of a parton which 
subsequently decays into the quarkonium state and  other partons.
This process is described by a fragmentation function $D(z,\mu)$, where 
$z$ is  the longitudinal momentum fraction of the quarkonium state and
$\mu$ is a factorization scale.  

The NRQCD factorization formalism \cite{B-B-L}
can be used to factor the fragmentation functions $D(z,\mu)$
for quarkonium into NRQCD matrix elements, which can be regarded as
phenomenological parameters, and short-distance factors, which
depend on $z$ and are calculable in perturbation theory.
Most of the phenomenologically relevant short-distance factors
begin at order $\alpha_s^2$ and have been calculated to leading order.
However there is one short-distance factor that begins at order $\alpha_s$.
It is the color-octet $^3S_1$ term in the gluon fragmentation function,
whose NRQCD matrix element is denoted  $\langle O_8(^3S_1) \rangle$.
This term is of particular phenomenological importance.  
Braaten and Yuan showed that it must be included in 
the gluon fragmentation function for triplet P-wave states in order to avoid 
an infrared divergence in the short-distance coefficient of the
color-singlet matrix element $\langle O_1(^3P_J) \rangle$ \cite{B-Y:P}.
Braaten and Fleming argued that the $\langle O_8(^3S_1) \rangle$ term is also 
phenomenologically necessary in the gluon fragmentation function
for spin-triplet S-wave states in order to explain the production rate of 
direct $J/\psi$ and $\psi'$ at large $p_T$ at the Tevatron \cite{B-F}.
This led to the remarkable prediction by Cho and Wise that 
$J/\psi$ and $\psi'$ at large $p_T$ should be transversely polarized
\cite{C-W}.

In the earliest calculations of fragmentation functions for
heavy quarkonium \cite{B-Y:S1,B-C-Y:cS,B-C-Y:Bc},
the short-distance factors were deduced by comparing the cross sections 
for quarkonium production with the form predicted by the factorization
theorems for inclusive single-hadron production.
However the fragmentation functions can also be defined formally
as matrix elements of bilocal operators in a light-cone gauge 
\cite{C-F-P} or, more generally, as
matrix elements of non-local gauge-invariant operators \cite{C-S}.
The gauge-invariant definition of Collins and Soper 
was first applied to calculations 
of the fragmentation functions for heavy quarkonium by Ma \cite{Ma-1}.
The definition is particularly convenient for carrying out 
calculations beyond leading order in $\alpha_s$.
It was used by Ma to calculate the short-distance factor of
the color-octet $^3S_1$ term in the gluon fragmentation function
to next-to-leading order in $\alpha_s$ \cite{Ma-2}.

In this paper, we calculate
the short-distance coefficients for the color-octet $^3S_1$ term
in the fragmentation function for a gluon to split into
heavy quarkonium states to order $\alpha_s^2$.
We use the gauge-invariant definition of the fragmentation function
given by Collins and Soper\cite{C-S},
and we remove ultraviolet divergences using the 
$\overline{\rm MS}$ renormalization procedure.
Our result for the longitudinal term in the fragmentation function
agrees with a previous calculation by Beneke and Rothstein \cite{B-R}.
Our result for the next-to-leading order correction in the transverse 
term disagrees with a previous calculation by Ma \cite{Ma-2}.

\section{Gauge-invariant Definition}

The fragmentation function $D_{g \to H}(z,\mu)$
gives the probability that a gluon produced in a hard-scattering process
involving momentum transfer of order $\mu$ decays into a hadron $H$
carrying a fraction $z$ of the gluon's longitudinal momentum.
This function can be defined in terms of the matrix element of a bilocal 
operator involving two gluon field strengths in a light-cone gauge
\cite{C-F-P}.  In Ref. \cite{C-S},
Collins and Soper introduced a gauge-invariant definition of the gluon 
fragmentation function that involves the matrix element 
of a nonlocal operator consisting of two gluon field strengths and 
eikonal operators.  One advantage of this definition is that it 
avoids subtleties associated with products of singular distributions.  
The gauge-invariant definition is also advantageous for explicit 
perturbative calculations, because it allows the calculation of 
radiative corrections to be simplified by using Feynman gauge.   

The gauge-invariant definition of Collins and Soper is
\begin{eqnarray}
D_{g \to H}(z,\mu) &=&
{(-g_{\mu \nu})z^{N-2} \over 16\pi(N-1) k^+}
\int_{-\infty}^{+\infty} dx^- e^{-i k^+ x^-} 
\nonumber\\
&&
\langle 0 | G^{+\mu}_c(0)
{\cal E}^\dagger(0^-)_{cb} \; {\cal P}_{H(z k^+,0_\perp)} \;
{\cal E}(x^-)_{ba} G^{+ \nu}_a(0^+,x^-,0_\perp) | 0 \rangle\;.
\label{D-def}
\end{eqnarray}
The operator ${\cal E}(x^-)$ in (\ref{D-def}) is an eikonal operator
that involves a path-ordered exponential of gluon field operators along
a light-like path:
\begin{equation}
{\cal E}(x^-)_{ba} \;=\; {\rm P} \exp 
\left[ +i g \int_{x^-}^\infty dz^- A^+(0^+,z^-,0_\perp) \right]_{ba},
\end{equation}
where $A^\mu(x)$ is the matrix-valued gluon field in the adjoint
representation:  $[ A^\mu(x) ]_{ac} = if^{abc} A_b^\mu(x)$.
The operator ${\cal P}_{H(p^+,p_\perp)}$ in (\ref{D-def})
is a projection onto states that in the asymptotic future contain 
a hadron $H$ with momentum $p = (p^+,p^-=(m_H^2+ p_\perp^2)/p^+,p_\perp)$, 
where $m_H$ is the mass of the hadron.
In the definition (\ref{D-def}), the hard-scattering scale $\mu$
can be identified with the renormalization scale of the nonlocal operator.
In perturbative calculations, it is convenient to use dimensional 
regularization to regularize ultraviolet divergences.  
The prefactor in the definition (\ref{D-def})
has therefore been expressed as a function of the number 
of spatial dimensions $N=3-2\epsilon$.
\begin{figure}
\begin{center}
\epsfig{file=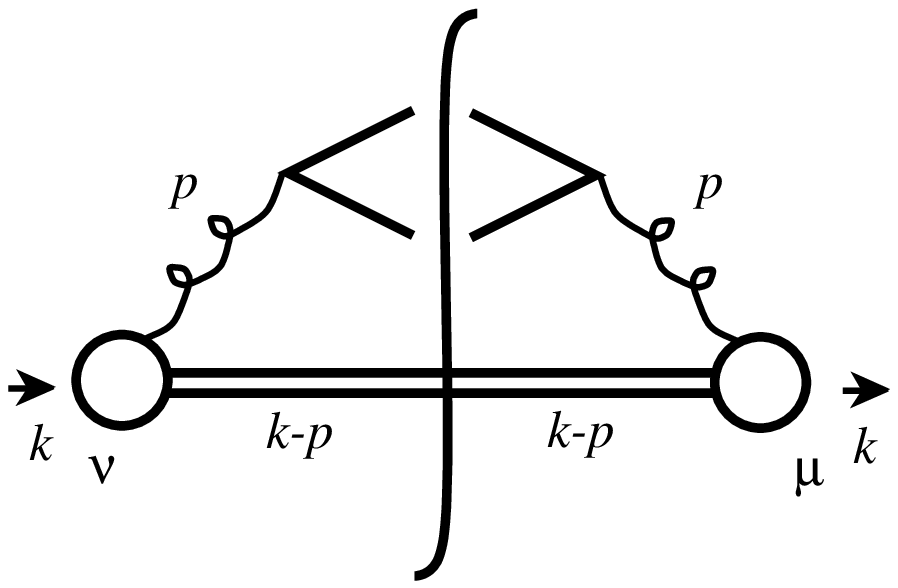, height=30 ex}
\end{center}
\caption{Leading order Feynman diagram for $g \to Q\overline{Q}$.}
\label{2body0}
\end{figure}

For any state $H$ that can be defined in perturbation theory,
the definition (\ref{D-def}) can be used to calculate the fragmentation 
function $D_{g\to H}(z,\mu)$ as a power series in $\alpha_s$.
A convenient set of Feynman rules for this perturbative expansion 
is given in Ref.\cite{C-S}.
If the state consists of a $Q \overline{Q}$ pair
with invariant mass $p^2$,
the lowest order diagram is shown in Fig.~\ref{2body0}.
The circles connected by the double pair of lines represent the 
nonlocal operator consisting of the gluon field strengths and 
the eikonal operators.
The momentum $k = (k^+,k^-,k_\perp)$ flows into the circle on the left
and out the circle on the right. 
The cutting line represents the projection onto states that in the 
asymptotic future include a $Q \overline{Q}$ pair with total momentum
$p = (z k^+, p^2/(z k^+), 0_\perp)$.
If the heavy quark has relative momentum ${\bf q}$ in the $Q \overline{Q}$
rest frame, the invariant mass is $p^2 = 4 (m_Q^2 + {\bf q}^2)$.

The fragmentation of a gluon into heavy quarkonium state $H$
involves many momentum scales, 
ranging from the hard-scattering scale $\mu$,
which we will assume to be larger than $m_Q$, to momenta much smaller than 
$m_Q$ where nonperturbative effects are large. 
The {\it NRQCD factorization formalism} 
allows the systematic separation of momentum scales of order $m_Q$ 
and larger from  scales of order $m_Q v$ or smaller, 
where $v$ is the typical relative velocity of the heavy quark in the hadron.
The factorization formula has the form
\begin{eqnarray}
D_{g \to H}(z,\mu) \;=\; 
\sum_{mn} d_{mn}(z,\mu) \langle {\cal O}_{nm}^H \rangle\;.
\label{NRQCD-fact}
\end{eqnarray}
The {\it short-distance coefficients} $d_{mn}(z,\mu)$ are independent of the 
quarkonium state $H$ and can be 
calculated as a perturbation series in $\alpha_s(m_Q)$.
All long-distance effects are factored into the {\it NRQCD matrix elements} 
$\langle {\cal O}_{nm}^H \rangle$, which can be expressed as matrix elements 
in an effective field theory.  They have the general form
\begin{eqnarray}
\langle {\cal O}_{nm}^H \rangle &=&
\langle 0 | \chi^\dagger {\cal K}_n \psi \; {\cal P}_H \; 
	\psi^\dagger {\cal K}_m \chi | 0 \rangle\;,
\end{eqnarray}
where ${\cal K}_m$ and ${\cal K}_n$ are constructed out of color matrices,
spin matrices, and covariant derivatives and the operator 
${\cal P}_H$ projects onto states that in the asymptotic future 
contain a quarkonium state $H$.
The NRQCD matrix elements are nonperturbative but they are universal, 
with the same matrix elements describing  
inclusive production in other high energy processes.

The threshold expansion method \cite{B-C:TE,B-C:TEdr} is 
a general prescription for determining the short-distance coefficients 
$d_{nm}(z,\mu)$ in the NRQCD factorization formula (\ref{NRQCD-fact}).
The diagrams for the fragmentation function of a $Q \overline{Q}$ pair
are computed in perturbation theory, except that the $Q \overline{Q}$ pair
is allowed to be in a different state on the two sides of the final-state
cut.  To the left of the cutting line, the $Q$ and $\overline{Q}$ have relative
momentum ${\bf q}$ in the $Q \overline{Q}$ rest frame and color and spin 
states specified by Pauli spinors $\eta$ and $\xi$.
To the right of the cutting line, the $Q$ and $\overline{Q}$ have relative
momentum ${\bf q}'$ and color and spin states specified by $\eta'$ and $\xi'$.
After expanding around the threshold ${\bf q} = {\bf q}' = 0$,
the resulting expression for the diagrams has the form
\begin{eqnarray}
\sum_{mn} d_{mn}(z,\mu) \; {\eta'}^\dagger \kappa_n \xi' 
			\; \xi^\dagger \kappa_m \eta,
\end{eqnarray}
where the color and spin matrices $\kappa_m$ and $\kappa_n$ are  
polynomials in the relative momenta ${\bf q}$ and ${\bf q}'$.
The short-distance coefficients $d_{nm}(z,\mu)$ in the factorization formula 
(\ref{NRQCD-fact}) can then be read off from this expression.
For example, if the sum of the diagrams includes the 
color-octet $^3S_1$ terms
\begin{eqnarray}
N m_Q \left[ d_T(z) ( \delta^{ij} - \hat z^i \hat z^j) 
	+ d_L(z) \hat z^i \hat z^j  \right] 
{\eta'}^\dagger \sigma^j T^a \xi' \; \xi^\dagger \sigma^i T^a \eta,
\label{d-match}
\end{eqnarray}
then the fragmentation function includes the terms
\begin{eqnarray}
D_{g \to H}(z) &=&
{N \over 4 m_Q} \left[ d_T(z) ( \delta^{ij} - \hat z^i \hat z^j) 
	+ d_L(z) \hat z^i \hat z^j  \right] 
\; \langle 0 | \chi^\dagger \sigma^j T^a \psi \; {\cal P}_H \; 
	\psi^\dagger \sigma^i T^a \chi | 0 \rangle.
\label{D-TL}
\end{eqnarray}
If we sum over the spin states of $H$, the matrix element 
in (\ref{D-TL}) is proportional to  $\delta^{ij}$ 
and (\ref{D-TL}) reduces to
\begin{eqnarray}
D_{g \to H}(z) \;=\;
\left[ (N-1) d_T(z) + d_L(z)  \right] \langle {\cal O}_8^H(^3S_1) \rangle,
\label{D-sum}
\end{eqnarray}
where $\langle {\cal O}_8^H(^3S_1) \rangle$ is the color-octet $^3S_1$ 
matrix element  defined in Ref. \cite{B-B-L}:
\begin{eqnarray}
\langle {\cal O}_8^H(^3S_1) \rangle\;=\;
{1 \over 4 m_Q} 
\langle 0 | \chi^\dagger \sigma^i T^a \psi \; {\cal P}_H \; 
	\psi^\dagger \sigma^i T^a \chi | 0 \rangle.
\end{eqnarray}
The factor of $1/( 4 m_Q)$ accounts for the relativistic normalization 
of the projection operator used in Refs. \cite{B-C:TE,B-C:TEdr}.

\section{Leading order}

The only Feynman diagram of order $\alpha_s$ for the fragmentation 
process $g \to Q \overline{Q}$ is shown
in Fig.~\ref{2body0}.  Using the Feynman rules of Ref. \cite{C-S},
the expression for the diagram can be easily written down in terms 
of spinors $u$ and $v$ that describe the color, spin, and relative momentum 
states of the $Q$ and $\overline{Q}$.  We can allow for the 
$Q$ and $\overline{Q}$ on the right side of the final-state cut 
to have different color and spin states and different relative 
momentum ${\bf q}'$ by replacing their spinors by $u'$ and $v'$.
The resulting expression is
\begin{eqnarray}
\frac{\pi\alpha_s\mu^{2\epsilon}}{2(N-1)(p^2)^2}
\; \delta(1-z)\;
\left( - g^{\alpha \beta} - {p^2 \over (k \cdot n)^2}n^\alpha n^\beta  \right)
{\bar v}' \gamma_\beta T^a u' \; \bar u \gamma_\alpha T^a v.
\label{d1-rel}
\end{eqnarray}
Setting ${\bf q} = {\bf q}' = 0$, we can replace the Dirac spinors with 
Pauli spinors by substituting
\begin{eqnarray}
\bar u \gamma_\alpha T^a v \;=\;
2 m_Q \; L_{\alpha i} \; \xi^\dagger \sigma^i T^a \eta,
\end{eqnarray}
where $L_{\alpha i}$ is a boost matrix that satisfies the identities
\begin{eqnarray}
-g^{\alpha \beta} L_{\alpha i} L_{\beta j} &=& \delta^{i j},
\\
n^\alpha L_{\alpha i} &=& {p \cdot n \over \sqrt{p^2}} \hat z^i.
\end{eqnarray}
The expression (\ref{d1-rel}) then reduces to 
\begin{eqnarray}
{\pi \alpha_s \mu^{2\epsilon} \over 8 (N-1) m_Q^2}
\; \delta(1-z)\; ( \delta^{ij} - \hat z^i \hat z^j)
{\eta'}^\dagger \sigma^j T^a \xi' \; \xi^\dagger \sigma^i T^a \eta.
\label{d1-nr}
\end{eqnarray}
Comparing with (\ref{d-match}),
we can read off the order-$\alpha_s$ terms in the short-distance functions
$d_T(z)$ and $d_L(z)$ defined in (\ref{D-TL}):
\begin{eqnarray}
d_T^{(\rm LO)}(z) &=&
\frac{\pi\alpha_s\mu^{2\epsilon}}{8N(N-1)m_Q^3}
\;\delta(1-z),
\label{d1-T}
\\
d_L^{(\rm LO)}(z) &=& 0.
\label{d1-L}\end{eqnarray}
The dependence on the number of spatial dimensions $N$ 
agrees with that in Ref. \cite{B-C:TEdr}.
\section{Virtual Corrections}

\begin{figure}
\begin{center}
\epsfig{file=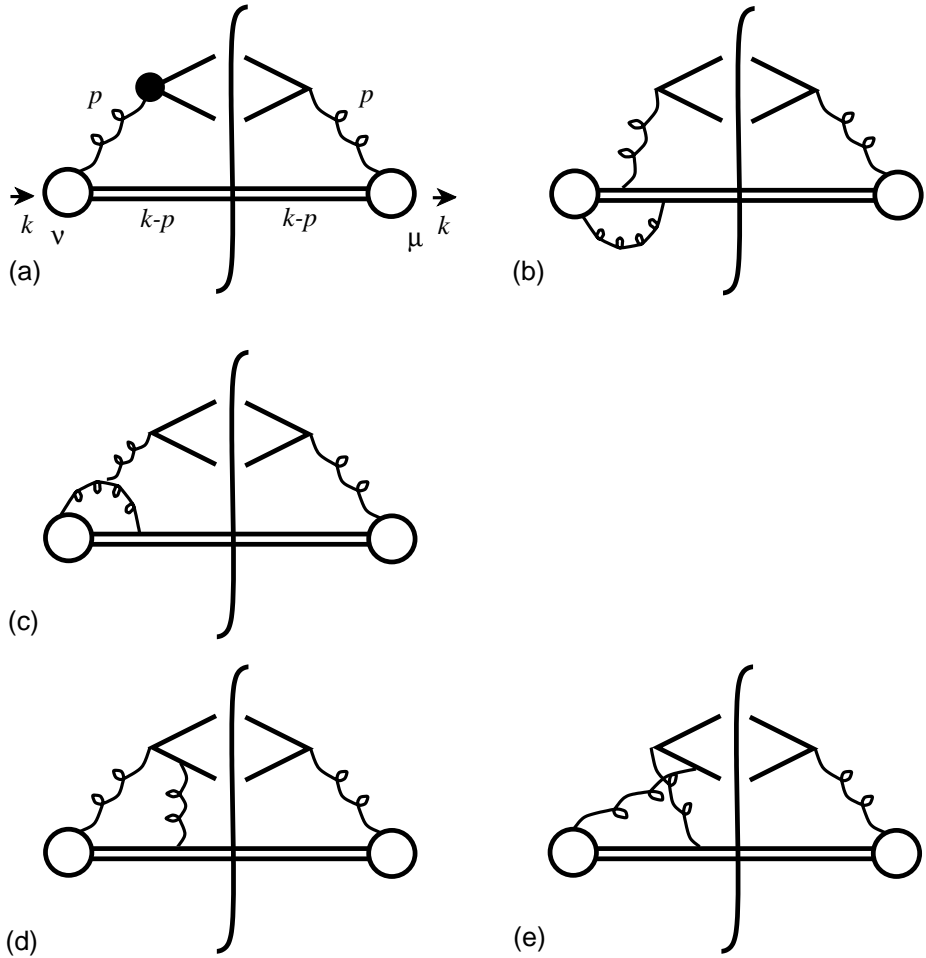, height=70ex}
\end{center}
\caption{The Feynman diagrams of order $\alpha_s^2$ for $g \to Q\overline{Q}$
with $Q\overline{Q}$ final states.  There are additional contributions 
from the complex-conjugate diagrams.}
\label{2bodya}
\end{figure}

\begin{figure}
\begin{center}
\epsfig{file=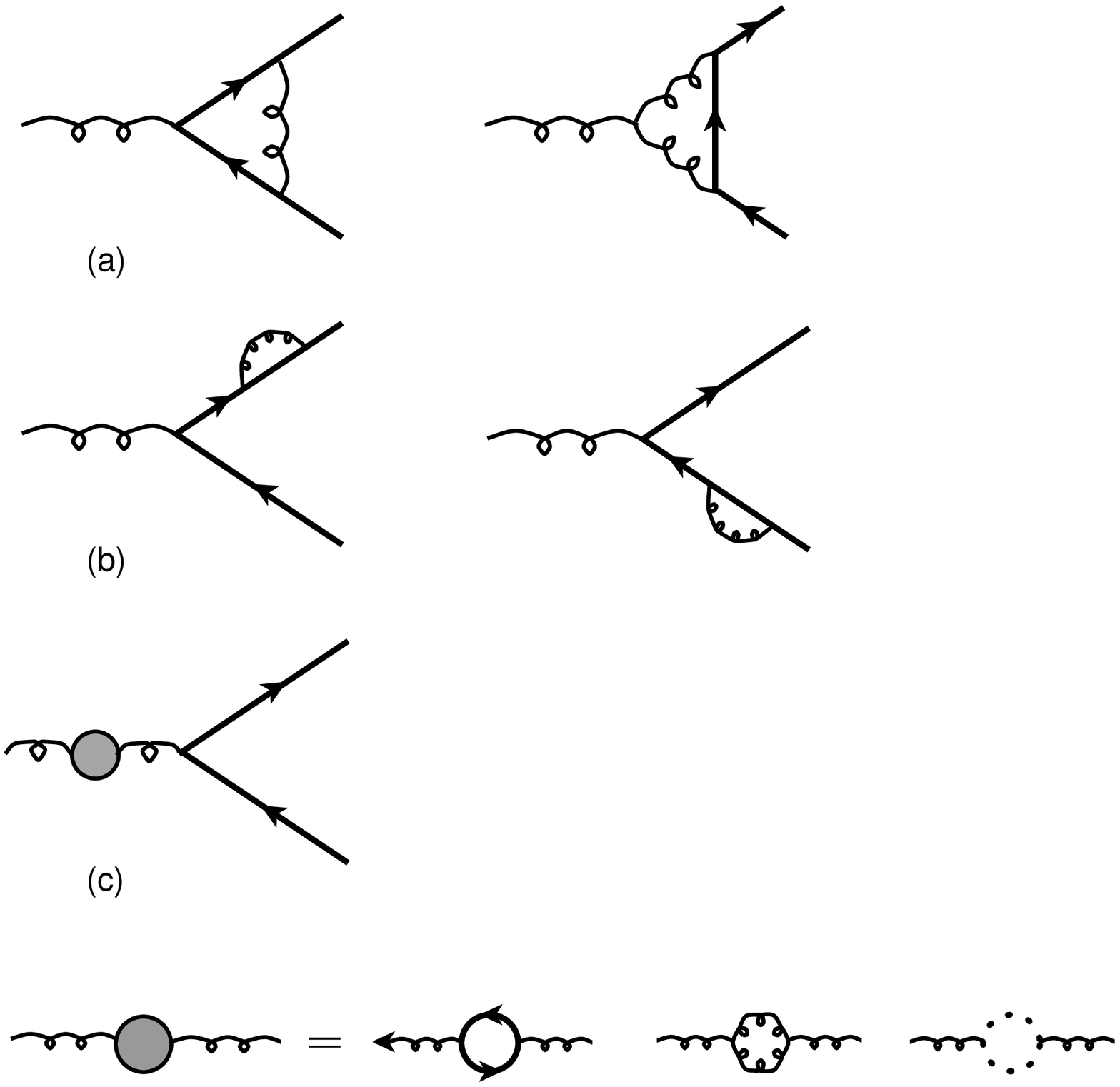, height=52ex}
\end{center}
\caption{One loop correction diagrams for
$g^*\to Q\overline{Q}$.
        }
\label{2onlya}
\end{figure}

The Feynman diagrams for the fragmentation function 
for $g \to Q\overline{Q}$ at order $\alpha_s^2$
consist of virtual corrections, for which the final state is $Q\overline{Q}$,
and real-gluon corrections, for which the final state is $Q\overline{Q}g$.
The diagrams with virtual-gluon corrections to the left of the cutting line
are shown in Fig.~\ref{2bodya}.  The black blob in  Fig.~\ref{2bodya}(a)
includes the vertex corrections and propagator corrections shown in 
Fig.~\ref{2onlya}.

We calculate the diagrams using Feynman gauge.
In this case, the diagram Fig.~\ref{2bodya}(b) vanishes,
because the gluon attaching to the eikonal line gives a factor of $n^\mu$.
When contracted with the factor $k \cdot n g^{\nu \beta} - p^\nu  n^\beta$
from the circle on the left side of the cut, it gives a factor
$(k - p) \cdot n = 0$.  In the limit ${\bf q} = {\bf q}' = 0$,
all the other diagrams can be reduced to the leading order diagram 
in Fig.~\ref{2body0} times a multiplicative factor.
For the diagrams in Fig.~\ref{2bodya}(a), this follows from the Dirac equation 
for the heavy-quark spinors. For the remaining diagrams, 
we must also use the fact that the contraction of $n^\mu$ with the 
right side of the diagram vanishes.

The virtual corrections contribute only to the transverse short-distance 
function $d_T(z)$ defined in (\ref{D-TL}).  We will express the
various contributions in the form of the leading-order result (\ref{d1-T})
times a multiplicative factor.  
The sum of the diagrams in  Fig.~\ref{2bodya}(a), together with their 
complex conjugates, is
\begin{eqnarray}
d_T^{(2a)}(z) &=& d_T^{(\rm LO)}(z) \times
2\;{\rm Re}\bigg[\Lambda + \Pi + (Z_Q-1) \bigg],
\label{d-2a}
\end{eqnarray}
where $\Lambda$ is the vertex correction factor from the diagrams 
in Fig.~\ref{2onlya}(b), 
$\Pi$ is the propagator correction factor for a gluon with invariant 
mass $4m_Q^2$ from the diagrams in Fig.~\ref{2onlya}(d),
and $Z_Q-1$ comes from the wavefunction renormalization factors $Z_Q^{1/2}$ 
for the heavy quark from the diagrams in Fig.~\ref{2onlya}(c).
These correction factors are given by
\begin{eqnarray}
\Lambda &=& 
{\alpha_s \over \pi} \left( {\pi \mu^2 \over m_Q^2} \right)^\epsilon
\left[ {13 \over 12} {\Gamma(1+\epsilon) \over \epsilon_{UV}} 
	- {1 \over 12} {\Gamma(1+\epsilon) \over \epsilon_{IR}}
	+ {13 \over 6} + 3 \ln 2 - i {\pi \over 2} \right],
\\
Z_Q-1 &=& 
{\alpha_s \over \pi}  \left( {\pi \mu^2 \over m_Q^2} \right)^\epsilon
\left[ - {1 \over 3} {\Gamma(1+\epsilon) \over \epsilon_{UV}}
	- {2 \over 3} {\Gamma(1+\epsilon) \over \epsilon_{IR}}
	 - {4 \over 3} - 2 \ln 2 \right],
\\
\Pi &=& 
{\alpha_s \over \pi}  \left( {\pi \mu^2 \over m_Q^2} e^{i \pi} \right)^\epsilon
\left[ {15-2n_f \over 12} {\Gamma(1+\epsilon) \over \epsilon_{UV}}
	+ {93-10 n_f \over 36} \right],
\end{eqnarray}
where $n_f$ is the number of light quark flavors.
The subscripts on the poles in $\epsilon$ indicate whether the divergences
are of ultraviolet or infrared origin.
The sum of the virtual corrections given in (\ref{d-2a}) is 
\begin{eqnarray}
d_T^{(2a)}(z) &=& d_T^{(\rm LO)}(z) \times
{\alpha_s \over \pi}  \left( {\pi \mu^2 \over m_Q^2} \right)^\epsilon
\Gamma(1+\epsilon) \left[  {12 - n_f \over  3 \epsilon_{UV}} 
	- {3 \over 2 \epsilon_{IR}}
	+ {123-10 n_f \over 18} + 2 \ln 2  \right],
\label{d-virt:2a}
\end{eqnarray}
Ma's result for these diagrams \cite{Ma-2} differs by an additive constant 
$-5.487$ inside the square brackets and by setting $N(N-1) \to 3(N-1)$ 
in $d_T^{(\rm LO)}(z)$.

The contribution from the diagrams 
in Fig.~\ref{2bodya}(c) with its complex conjugate is
\begin{eqnarray}
d_T^{(2c)}(z) &=& d_T^{(\rm LO)}(z)
\times 24 \pi \alpha_s \mu^{2\epsilon} \;
{\rm Re}\bigg[i \left(2 k \cdot n I_{ABC}-I_{AB}\right)\bigg],
\label{d-2c}
\end{eqnarray}
where the scalar integrals $I_{AB\cdots}$ are given in appendix A.
The contribution from the sum of the diagrams in Figs.~\ref{2bodya}(d) 
and \ref{2bodya}(e) with their complex conjugates is
\begin{eqnarray}
d_T^{(2d,e)}(z) &=&d_T^{(\rm LO)}(z) \times
96 \pi \alpha_s \mu^{2\epsilon} m_Q^2 \;
{\rm Re} \bigg[ i \left( k \cdot n I_{ABCD} -I_{ABD} \right) \bigg].
\label{d-2de}
\end{eqnarray}
The sum of the virtual corrections given in (\ref{d-2c}) 
and (\ref{d-2de}) is 
\begin{eqnarray}
d_T^{(2c,d,e)}(z) &=& d_T^{(\rm LO)}(z) \times
{\alpha_s \over \pi}  \left( {\pi \mu^2 \over m_Q^2} \right)^\epsilon
\Gamma(1+\epsilon)
\left[  {3(1-\epsilon) \over 2 \epsilon_{UV} \epsilon_{IR}} 
	+ {3 \over 2\epsilon_{UV}}
\right.
\nonumber \\ 
&& \left. \qquad \qquad \qquad \qquad \qquad
	+ 3 - {\pi^2 \over 2} 
	+ 6 \ln 2 + 6 \ln^2 2  \right].
\label{d-virt:2cde}
\end{eqnarray}
Ma's result for these diagrams  \cite{Ma-2} differs by an additive constant 
$-3 + {3 \over 2} \pi^2 - 6 \ln^2 2$
inside the square brackets and by setting $N(N-1) \to 3(N-1)$ 
in $d_T^{(\rm LO)}(z)$.

The total virtual correction is the sum of (\ref{d-virt:2a})
and (\ref{d-virt:2cde}):
\begin{eqnarray}
d_T^{\rm (virtual)}(z) &=& d_T^{(\rm LO)}(z) \times
{\alpha_s \over \pi}  \left( {\pi \mu^2 \over m_Q^2} \right)^\epsilon
\left[  {3(1-\epsilon) \over 2}{\Gamma(1+\epsilon) \over 
         \epsilon_{UV} \epsilon_{IR}} 
	+ \beta_0 {\Gamma(1+\epsilon) \over \epsilon_{UV}}
\right.
\nonumber \\ 
&& \left. \qquad \qquad \qquad \qquad \qquad
	+ {177-10n_f \over 18} - {\pi^2 \over 2} 
	+ 8 \ln 2 + 6 \ln^2 2  \right],
\label{d-virtual}
\end{eqnarray}
where $\beta_0=(33-2n_f)/6$.

\section{Real Gluon Corrections}
\begin{figure}
\begin{center}
\epsfig{file=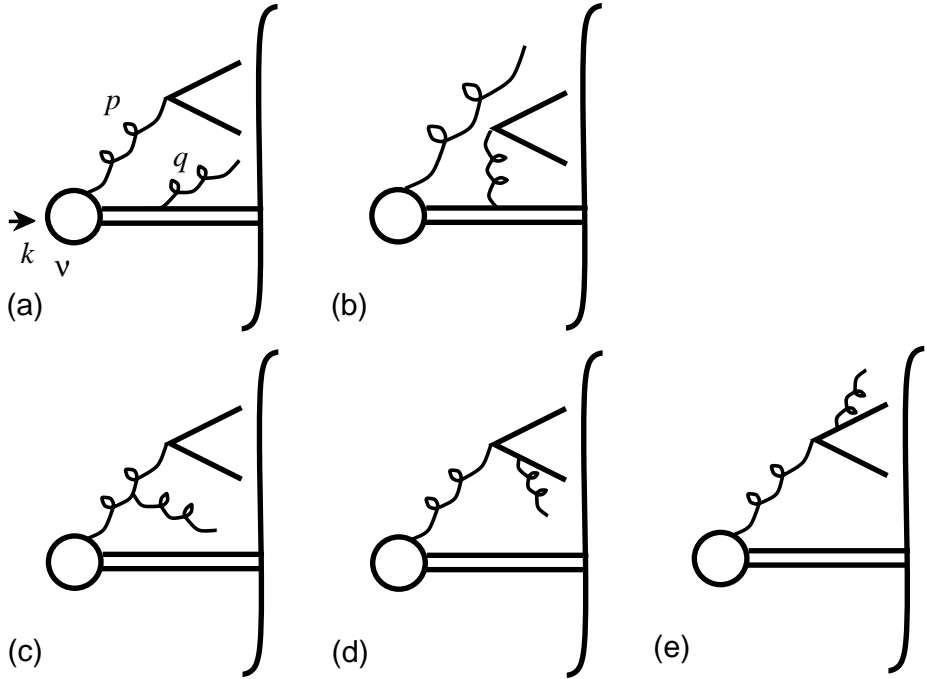, height=50ex}
\end{center}
\caption{
The Feynman diagrams of order $\alpha_s^2$ for $g \to Q\overline{Q}$
with $Q\overline{Q}g$ final states.  There are a total of 25 diagrams, 
but only the left halves of the diagrams are shown.}
\label{3body}
\end{figure}
The Feynman diagrams for the real-gluon corrections to the fragmentation 
function for $g\to Q\overline{Q}$ pair are shown in Fig.~\ref{3body}.
We draw the 5 left-half diagrams only, but they must be multiplied by
their complex conjugates to give a total of 25 diagrams.
Only diagrams \ref{3body}(a), \ref{3body}(b) and \ref{3body}(c)
contribute to the color-octet $^3S_1$ term, which reduces the total
number of diagrams to 9.
We calculate the diagrams using Feynman gauge.  
After a considerable amount of algebra, they reduce to
\begin{eqnarray}
d_T^{\rm  (real)}(z)&=&
\frac{\pi\alpha_s\mu^{2\epsilon}}
{8N(N-1)m_Q^3}
\times
\frac{3\alpha_s}{\pi\Gamma(1-\epsilon)}
\left(
\frac{\pi \mu^2}{m_Q^2}
\right)^\epsilon
\left(1-\frac{1}{z(1-z)}\right)^2
\int_{(1-z)/z}^\infty
dx
\frac{t^{1-\epsilon}}{x^2}\;,
\label{eq:epsIR}
\\
d_L^{\rm  (real)}(z)&=&
\frac{\pi\alpha_s\mu^{2\epsilon}}
{8Nm_Q^3}
\times
\frac{3\alpha_s}{\pi\Gamma(1-\epsilon)}
\left(
\frac{\pi \mu^2}{m_Q^2}
\right)^\epsilon
\left(\frac{1-z}{z}\right)^2
\int_{(1-z)/z}^\infty
dx
\frac{t^{-\epsilon}}{x^2}\;,
\end{eqnarray}
where $t=(1-z)(zx+z-1)$, $x=2q\cdot p/p^2$, 
$q$ is the final-state gluon momentum, and $p$ is the 
$Q\overline{Q}$ momentum. 
Integration over $x$ can be done using the following identities:
\begin{eqnarray}
\int_{(1-z)/z}^\infty dx\; \frac{t^{1-\epsilon}}{x^2}
&=&
z(1-z)^{1-2\epsilon}\;
\frac{1-\epsilon}{\epsilon_{UV}}
\; B(1+\epsilon, 1-\epsilon)\;,
\label{eq:epsUV}
\\
\int_{(1-z)/z}^\infty dx\; \frac{t^{-\epsilon}}{x^2}
&=&
z(1-z)^{-1-2\epsilon}\;
B(1+\epsilon, 1-\epsilon)\;.
\end{eqnarray}
The subscript on $\epsilon$ in (\ref{eq:epsUV}) indicates that the pole
has an ultraviolet origin. 
In (\ref{eq:epsIR}), there is also an infrared divergence
associated with the limit $z\to 1$.
It can be made explicit by using the expansion:
\begin{eqnarray}
\frac{1}{(1-z)^{1+2\epsilon}}
&=&
-\frac{1}{2\epsilon_{IR}}
\delta(1-z)
+\frac{1}{(1-z)_+}
-2\epsilon\left(\frac{\ln(1-z)}{1-z}\right)_+
+O(\epsilon^2)\;.
\end{eqnarray}
The final results for the real-gluon corrections are
\begin{eqnarray}
d_T^{\rm (real)}(z)
&=&
\frac{\pi\alpha_s\mu^{2\epsilon}}
{8N(N-1)m_Q^3}
\times
\frac{\alpha_s}{\pi}
\;
\left(\frac{\pi\mu^2}{m_Q^2}\right)^\epsilon \Gamma(1+\epsilon)
\;\bigg[
-\frac{3(1-\epsilon)}{2\epsilon_{UV}\epsilon_{IR}}
\delta(1-z)
\nonumber\\
&&
+\frac{3(1-\epsilon)}{\epsilon_{UV}}
\left(
\frac{z}{(1-z)_+}+\frac{1-z}{z}+z(1-z)
\right)
\nonumber\\
&&
~~  -\frac{6}{z}\left(\frac{\ln(1-z)}{1-z}\right)_+
+6(2-z+z^2)\ln(1-z)
\bigg]\;,
\label{d-real}
\\
d_L^{\rm (real)}(z)
&=&
\frac{\pi\alpha_s}
{8Nm_Q^3}
\times
\frac{3\alpha_s}{\pi}
\frac{1-z}{z}\;.
\label{dL-real}
\end{eqnarray}
Note that the double-pole term proportional to
$1/(\epsilon_{UV}\epsilon_{IR})$ in (\ref{d-real}) exactly cancels 
its counter part in the  virtual correction (\ref{d-virtual}).
The sum of (\ref{d-virtual}) and (\ref{d-real})
is free of infrared divergences.
Ma's result for the real-gluon corrections  \cite{Ma-2} is completely 
different from the sum of (\ref{d-real}) and (\ref{dL-real}).
In particular, he found that the $[\ln(1-z)/(1-z)]_+$ terms
cancelled.  In the way we organized the calculation, 
there is no possibility of such a cancellation.

\section{Renormalization}
The sum of the order-$\alpha_s^2$ corrections to 
$d_T(z)$ in 
(\ref{d-virtual}) and (\ref{d-real})
still contains ultraviolet divergences in
the form of single poles in $\epsilon$.
These divergences are cancelled by the renormalization of the 
coupling constant $\alpha_s$ in the leading-order expression (\ref{d1-L}) and
by the renormalization of the nonlocal operator in (\ref{D-def}).
The renormalization of $\alpha_s$ in the $\overline{\rm MS}$ scheme
can be carried out by making the following substitution in (\ref{d1-L}):
\begin{eqnarray}
\alpha_s\to
\alpha_s
\left[
1-\frac{\alpha_s}{2\pi}\;
\beta_0 \left(4\pi e^{-\gamma}\right)^\epsilon \frac{1}{\epsilon_{UV}}
\right]\,,
\end{eqnarray}
where $\beta_0=(33-2n_f)/6$.
The operator renormalization in the $\overline{\rm MS}$ scheme
can be carried out by making the following substitution
\begin{eqnarray}
d^{(\rm LO)}_T(z)
\to
d^{(\rm LO)}_T(z)
-
\frac{\alpha_s}{2\pi}
\left(4\pi e^{-\gamma}\right)^\epsilon \frac{1}{\epsilon_{UV}}
\int_z^1 \frac{dy}{y}\; P_{gg}(y)\; d^{(\rm LO)}_T(z/y) \,,
\end{eqnarray}
where $P_{gg}(y)$ is the gluon splitting function:
\begin{eqnarray}
P_{gg}(z)=
6\left[
\frac{z}{(1-z)_+}+\frac{1-z}{z}+z(1-z)+\frac{\beta_0}{6}\;\delta(1-z)
\right]\;.
\end{eqnarray}
The sum of the two contributions of order $\alpha_s^2$ from
renormalization is
\begin{eqnarray}
d^{({\rm ren})}_T(z)
&=&
\frac{\pi\alpha_s\mu^{2\epsilon}}{8N(N-1)m_Q^3}
\nonumber\\
&\times&
\frac{\alpha_s}{\pi}
\left(4\pi e^{-\gamma}\right)^\epsilon 
\;
\frac{(-3)}{\epsilon_{UV}}
\left[
\frac{z}{(1-z)_+}+\frac{1-z}{z}+z(1-z)+\frac{\beta_0}{3}\;\delta(1-z)
\right]\;.
\label{d-ren}
\end{eqnarray}
This cancels the single ultraviolet poles in the sum of (\ref{d-virtual}) 
and (\ref{d-real}).
Our final result for the transverse fragmentation function is obtained by
adding the order-$\alpha_s^2$ corrections to $d_T(z)$ from
(\ref{d-virtual}), (\ref{d-real}), and (\ref{d-ren}) 
and taking $\epsilon\to 0$:
\begin{eqnarray}
d_T(z,\mu)
=
\frac{\pi\alpha_s(\mu)}{48m_Q^3}
\;
&\bigg\{&
\;
\delta(1-z)+\frac{\alpha_s(\mu)}{\pi}
\bigg[
A(\mu)\delta(1-z)
+\left(
 \ln\frac{\mu}{2m_Q}-\frac{1}{2}
 \right)P_{gg}(z)
\nonumber\\
&&+6(2-z+z^2)\ln(1-z)
  -\frac{6}{z}\left(\frac{\ln(1-z)}{1-z}\right)_+
\;
\bigg]
\bigg\}
\;,
\label{dT-final}
\end{eqnarray}
where the coefficient $A(\mu)$ is
\begin{eqnarray}
A(\mu)=
\beta_0\left(\ln\frac{\mu}{2m_Q}+\frac{13}{6}\right)
+\frac{2}{3}-\frac{\pi^2}{2}+8\ln 2+6\ln^2 2
\;.
\end{eqnarray}
The result (\ref{dT-final}) disagrees with the final result 
obtained by Ma \cite{Ma-2}.
Our final result for the longitudinal fragmentation function is 
obtained by setting $\epsilon \to 0$ in (\ref{dL-real}):
\begin{eqnarray}
d_L(z,\mu)
=\frac{\alpha_s^2(\mu)}{8m_Q^3}\;\frac{1-z}{z} \,.
\label{dL-final}
\end{eqnarray}
This agrees with the result of Beneke and Rothstein\cite{B-R}.
The fragmentation probabilities obtained by integrating $d_T(z)$ 
and $d_L(z)$ diverge, because these functions behave like $1/z$
as $z\to0$.
The higher moments of the fragmentation functions however are well-defined.

\section{Discussion}

The color-octet $^3S_1$ term in the fragmentation function is important 
for calculating the production at large $p_T$ of spin-singlet 
S-wave states, like the $J/\psi$, and spin-triplet P-wave states, 
like the $\chi_{cJ}$.  We can deduce the fragmentation functions 
for each of their spin states by using the approximate spin symmetry
of NRQCD to simplify the expression (\ref{D-TL}).  The functions
$d_T(z)$ in (\ref{dT-final}) and $d_L(z)$  (\ref{dL-final})
give the fragmentation functions for the transverse and longitudinal
spin states of the $J/\psi$, respectively: 
\begin{eqnarray}
D_{g \to J/\psi(\pm 1)}(z) &=& d_T(z) 
			\langle {\cal O}^{J/\psi}_8(^3S_1) \rangle \,,
\\
D_{g \to J/\psi(0)}(z) &=& d_L(z) 
			\langle {\cal O}^{J/\psi}_8(^3S_1) \rangle \,.
\end{eqnarray}
The sum over spin states is 
\begin{eqnarray}
D_{g \to J/\psi}(z) &=& \left[ 2 d_T(z) + d_L(z) \right]
			\langle {\cal O}^{J/\psi}_8(^3S_1) \rangle \,.
\end{eqnarray}
The fragmentation functions for each of the spin states of the 
$\chi_{cJ}$ are \cite{C-W-T}
\begin{eqnarray}
D_{g \to \chi_{c0}}(z) &=& \left[ 2 d_T(z) + d_L(z) \right]
			\langle {\cal O}_8^{\chi_{c0}}(^3S_1) \rangle \,,
\\
D_{g \to \chi_{c1}(0)}(z) &=& 3 d_T(z) 
			\langle {\cal O}_8^{\chi_{c0}}(^3S_1) \rangle \,,
\\
D_{g \to \chi_{c1}(\pm 1)}(z) &=& {3 \over 2} \left[ d_T(z) + d_L(z) \right]
			\langle {\cal O}_8^{\chi_{c0}}(^3S_1) \rangle \,,
\\
D_{g \to \chi_{c2}(0)}(z) &=&  \left[ d_T(z) +2  d_L(z) \right]
			\langle {\cal O}_8^{\chi_{c0}}(^3S_1) \rangle \,,
\\
D_{g \to \chi_{c2}(\pm 1)}(z) &=& {3 \over 2} \left[ d_T(z) + d_L(z) \right]
			\langle {\cal O}_8^{\chi_{c0}}(^3S_1) \rangle \,,
\\
D_{g \to \chi_{c2}(\pm 2)}(z) &=& 3 d_T(z) 
			\langle {\cal O}_8^{\chi_{c0}}(^3S_1) \rangle \,.
\end{eqnarray}
The sums over spin states are 
\begin{eqnarray}
D_{g \to \chi_{cJ}}(z) &=& (2J+1) \left[ 2 d_T(z) + d_L(z) \right]
			\langle {\cal O}^{\chi_{c0}}_8(^3S_1) \rangle \,.
\end{eqnarray}

In order to give accurate predictions for the production of quarkonium 
at large $p_T$, it is important to know the next-to-leading order 
correction to the color-octet $^3S_1$ term in the gluon fragmentation 
function.  Unfortunately our calculation of the short-distance coefficient
disagrees with the previous calculation by Ma.  An independent calculation 
of this important function is therefore essential. 

This work was supported in part by the U.S.
Department of Energy Division of High Energy Physics under
grant DE-FG02-91-ER40690, by the Alexander von Humboldt Foundation,
and by the Korea Institute for Advanced Study.
J.L. would like to thank the OSU theory group for
its hospitality during his stay in Columbus.
\appendix
\section{Integral Table}
In this appendix, we present the explicit values of the integrals
encountered in evaluating the 
virtual-gluon corrections.
These integrals have the form
\begin{equation}
I_{AB\cdots} \;=\; \int\frac{d^{N+1}l}{(2\pi)^{N+1}} \frac{1}{AB\cdots}\;,
\end{equation}
where the denominator $AB\cdots$ can be a product of 1, 2, 3, or 4 of
the following factors:
\begin{eqnarray}
A&=&l^2+i\epsilon,\\
B&=&(l-p)^2+i\epsilon=l^2-2l\cdot p+4 m_Q^2+i\epsilon,\\
C&=&(p-l)\cdot n+i\epsilon,\\
D&=&(l-p/2)^2-m_Q^2+i\epsilon=l^2-l\cdot p+i\epsilon.
\end{eqnarray}
The momentum $p$ is that of a $Q \overline{Q}$ pair with 
zero relative momentum ($p^2 = 4 m_Q^2$) and $n$ is light-like ($n^2 = 0$).  
The integrals $I_A$ and $I_B$ vanish in dimensional regularization.
By symmetry under $p \to l - p$, we have $I_{AD} = I_{BD}$.
Some of the integrals can be reduced to ones with fewer denominators 
by using the identity $A+B-2D=4 m_Q^2$:
\begin{eqnarray}
4 m_Q^2 I_{ABD} &=& 2(I_{AD}-I_{AB}),
\\
4 m_Q^2 I_{ABCD} &=& I_{ACD}+I_{BCD}-2I_{ABC}.
\end{eqnarray}
The independent integrals that need to be evaluated are therefore
\begin{eqnarray}
I_{AB} &=&
\frac{i}{(4\pi)^{2}}
\left( {\pi e^{i\pi} \over m_Q^2} \right)^\epsilon \;
{\Gamma(1+\epsilon) \Gamma^2(1-\epsilon) 
	\over \epsilon_{UV} \Gamma(2-2\epsilon)}\;,
\\
I_{AD}&=&
\frac{i}{(4\pi)^{2}}
\left( {4\pi \over m_Q^2} \right)^\epsilon \;
{\Gamma(1+\epsilon) \over \epsilon_{UV}(1 - 2\epsilon) } \;,
\\
I_{ABC}&=&
\frac{-i}{(4\pi)^{2}p\cdot n}
\left( {\pi e^{i\pi} \over m_Q^2} \right)^\epsilon \;
{\Gamma(1+\epsilon) \Gamma^2(1-\epsilon) 
	\over \epsilon_{UV} \epsilon_{IR} \Gamma(1-2\epsilon)}
\;,
\\
I_{ACD}&=&
\frac{+i}{(4\pi)^{2}p\cdot n}
\left( {4\pi \over m_Q^2} \right)^\epsilon \;
\frac{\Gamma(1+\epsilon)}{\epsilon_{UV}}\;
\left[ 2 \ln 2 + \epsilon \left( {\pi^2 \over 3} - 6 \ln^2 2 \right)
	+ O(\epsilon^2) \right] \;,
\\
I_{BCD}&=&
\frac{-i}{(4\pi)^{2}p\cdot n}
\left( {4\pi \over m_Q^2} \right)^\epsilon \;
{\Gamma(1+\epsilon) \over \epsilon_{UV} \epsilon_{IR}} \;.
\end{eqnarray}
The subscripts on the poles in $\epsilon$ indicate whether the divergences
are of ultraviolet or infrared origin.



\begin{references}

\bibitem{C-S}
J.C. Collins and D.E. Soper, Nucl. Phys. B {\bf 193}, 381 (1981);
	{\it ibid.} B {\bf 194}, 445 (1982).

\bibitem{B-Y:S1}
E. Braaten and T.C. Yuan, Phys. Rev. Lett.  {\bf 71}, 1673 (1993);
	Phys. Rev. D {\bf 52}, 6627 (1995).

\bibitem{B-B-L}
G.T. Bodwin, E. Braaten, and G.P. Lepage, Phys. Rev. D {\bf 51}, 1125 (1995).

\bibitem{B-Y:P}
E. Braaten and T.C. Yuan, Phys. Rev. D {\bf 50}, 3176 (1994);

\bibitem{B-F}
E. Braaten and S. Fleming, Phys. Rev. Lett. {\bf 74}, 3327 (1995).

\bibitem{C-W}
P. Cho and M.B. Wise, Phys. Lett. {\bf B346}, 129 (1995).


\bibitem{B-C-Y:cS}
E. Braaten, Kingman Cheung, and T.C. Yuan, 
	Phys. Rev. D {\bf 48}, 4230 (1993).

\bibitem{B-C-Y:Bc}
E. Braaten, Kingman Cheung, and T.C. Yuan, 
	Phys. Rev. D {\bf 48}, R5049 (1993).

\bibitem{C-F-P}
G. Curci, W. Furmanski, and R. Petronzio, Nucl. Phys. {\bf B175}, 27 (1980).

\bibitem{Ma-1}
J.P. Ma, Phys.Lett. {\bf B 332}, 398 (1994).

\bibitem{Ma-2}
J.P. Ma, Nucl. Phys. B {\bf 447}, 405 (1995).


\bibitem{B-R}
M. Beneke and I.Z. Rothstein, Phys. Lett. B {\bf 372}, 157 (1996);
	(E):{\it ibid.} B {\bf 389}, 769 (1996).

\bibitem{B-C:TE}
E. Braaten, Y.-Q. Chen, Phys. Rev. D {\bf 54}, 3216 (1996).

\bibitem{B-C:TEdr}
E. Braaten, Y.-Q. Chen, Phys. Rev. D {\bf 55}, 2693 (1997).

\bibitem{B-C:Pdecay}
E. Braaten, Y.-Q. Chen, Phys. Rev. D {\bf 55}, 7152 (1997).

\bibitem{C-W-T}
P. Cho, M.B. Wise, and S.P. Trivedi, Phys. Rev. D {\bf 51}, 2039 (1995). 
\end{references}
\end{document}